\documentclass[aps,pra,twocolumn,floatfix,superscriptaddress]{revtex4-2}
\usepackage{amsmath,bm} \usepackage{graphicx}
\usepackage[colorlinks=true,urlcolor=blue,breaklinks=true]{hyperref}

\graphicspath{{Figs_resub/}}

\begin{document}

\title{Self-sustained Josephson dynamics and self-trapping in
supersolids}

\author{Aitor Ala\~na} 
\affiliation{Department of Physics, University of the Basque Country
UPV/EHU, 48080 Bilbao, Spain} 
\affiliation{EHU Quantum Center,
University of the Basque Country UPV/EHU, 48940 Leioa, Biscay, Spain}
\author{Michele Modugno} 
\affiliation{Department of Physics, University of the Basque Country
UPV/EHU, 48080 Bilbao, Spain} 
\affiliation{IKERBASQUE, Basque
Foundation for Science, 48009 Bilbao, Spain} 
\affiliation{EHU Quantum
Center, University of the Basque Country UPV/EHU, 48940 Leioa, Biscay,
Spain} \author{Pablo Capuzzi} 
\affiliation{\hbox{Universidad de Buenos Aires, Facultad de Ciencias
Exactas y Naturales, Departamento de Física, 1428 Buenos Aires,
Argentina}} \affiliation{CONICET - Universidad de Buenos Aires,
Instituto de Física de Buenos Aires (IFIBA), 1428 Buenos Aires,
Argentina} \author{D. M. Jezek} 
\affiliation{CONICET - Universidad de Buenos Aires, Instituto de
Física de Buenos Aires (IFIBA), 1428 Buenos Aires, Argentina}

\date{\today}

\begin{abstract} We explore the self-sustained Josephson junction
dynamics in dipolar supersolids, predicting the possibility of
self-trapping alongside the experimentally observed Josephson
oscillations [Biagioni, G. \textit{et al.}, Nature \textbf{629}, 773
(2024)]. Using an asymmetric two-mode (ATM) model to describe a
triangular dipolar supersolid, validated through Gross-Pitaevskii
simulations, we demonstrate that the system's symmetry 
enables a consistent two-mode mapping despite the presence of
seven droplets. Hence, the associated  Hamiltonian allows us to
straightforwardly determine the self-trapping regime. 
Additionally, we show that bringing the system into rotation
preserves its ability to sustain the Josephson junction dynamics
across its full range, and we assess the robustness of the ATM model
under these conditions. 
We further find  that the off-axis droplets move in the radial
  direction during
the evolution in accordance with the size of the central
droplet. Such  movements do not interfere with the model
predictions.
\end{abstract}

\maketitle

\textit{Introduction.}  Bose-Einstein condensates (BECs) have proven
to be a versatile and tunable platform for realizing the Josephson
effect \cite{josephson1962}, a paradigmatic manifestation of
macroscopic quantum coherence \cite{pitaevskii2016}. The dynamics of
Josephson junctions, encompassing both oscillations of the population
imbalance around an equilibrium value and self-trapping, have been
extensively studied in BECs confined in double-well potentials, both
experimentally and theoretically
\cite{smerzi97,zapata1998,ragh99,giovanazzi2000,cataliotti2001,albeiz05,anan06,levy2007,jezek13a,adhikari2014,spagnolli2017,martinezg2018,mistakidis2024}.
Other configurations, where the junction is self-sustained without the need for
an external potential, have also been proposed \cite{abad11a,abad11b}.
Recently, the advent of dipolar supersolids
\cite{Ta19,Bo19,Ch19,boni12,Ch18,Na19,Ta21,No21,sohmen2021,Bl22} has
opened fascinating new perspectives in this respect. Notably,
self-sustained Josephson oscillations have been observed in a recent
experiment involving a quasi one-dimensional geometry with multiple
droplets \cite{Biagioni2024}. Additionally, a Josephson junction array
model has been used to describe another experiment that probed the
response of a dipolar supersolid to an interaction quench, leading to
the shattering of global phase coherence \cite{ilzhofer2021}.
However, the existence of self-trapping regimes in supersolids has not
yet been explored, and the sucess  of such a study  depends on the geometry of
the distribution of the droplets.

The study of Josephson dynamics in BECs confined to double-well
potentials has traditionally relied on two-mode models, which provide
a simplified description that effectively captures key phenomena such
as coherent oscillations and self-trapping
\cite{smerzi97,ragh99,anan06,juliadiaz2010,jezek13a,burchianti2017}.
These models most often assume symmetric configurations, where both
wells are identical \cite{albeiz05,mauro17}. Asymmetric two-mode (ATM)
models, which account for a nonzero equilibrium population imbalance,
have also been developed and shown to agree with experiments
\cite{cat14,ryu13}. In the case of supersolids, the Josephson junction
involves multiple interacting droplets, making theoretical
descriptions significantly more complex. For example, recent
experiments on dipolar supersolids required models with at least six
modes to capture the observed dynamics \cite{Biagioni2024}.  For such
$n$-mode models, since the macroscopic coordinates lie in a space of
$2n-2$ dimensions, multiple types of orbits exist,  making it
difficult to predict self-trapping oscillations which have large
differences in the imbalances with respect to the equilibrium one,
without imposing further conditions.  In Ref. \cite{jezek2013}
the dynamics of $n$-well condensates have been studied using both
Gross-Pitaevskii equations and $n$-mode models. There, it may be
seen that while in a double-well configuration a separatrix can
be drawn between the Josephson and self-trapping regimes, for
larger values of $n$, the orbits have very different behaviors
depending on the initial conditions.

The goal of this work is to select a system suitable to be
described by means of a two-mode Hamiltonian, and hence the
self-trapping regime can be determined straightforwardly. The
inherent symmetry of certain supersolid configurations, such as
triangular arrays of droplets, offers an opportunity to reduce
the complexity of these descriptions. Equivalent droplets in such
configurations enable a consistent two-mode mapping, even in the
presence of multiple droplets, paving the way for a deeper
understanding of the system's dynamics.

Supersolids also offer the intriguing possibility of extending
Josephson dynamics to rotating systems. In such cases, the evolution
of the condensate is influenced by spatial phase variations induced by
rotation \cite{rot20}, including the formation of quantized vortices
\cite{castin1999,modugno2003,Foster2010,ka11,gal20,an20,je23,poli23,aitor},
which have been experimentally observed in both regular superfluids
\cite{williams10} and supersolids \cite{casotti2024}. Additionally,
the dynamics may be affected by other effects unique to rotation
\cite{rot20,roccu20,Roccuzzo22}, potentially challenging its
stability. Hence, one could also expect that the almost free movement
of droplets could lead to the failure of the model.  As a result,
assessing the robustness of Josephson theory in rotating systems is of
significant importance. In light of this, in this work we investigate
self-sustained Josephson dynamics in dipolar supersolids, focusing on
the triangular supersolid as a test case. Using an ATM model validated
through Gross-Pitaevskii simulations, we demonstrate the existence of
self-trapping dynamics under specific initial conditions.  Moreover,
we show that the system retains its ability to sustain Josephson
dynamics when subjected to rotation, highlighting the robustness of
the ATM model despite the added complexity.

\begin{figure}[t] \includegraphics[width=0.7\columnwidth]{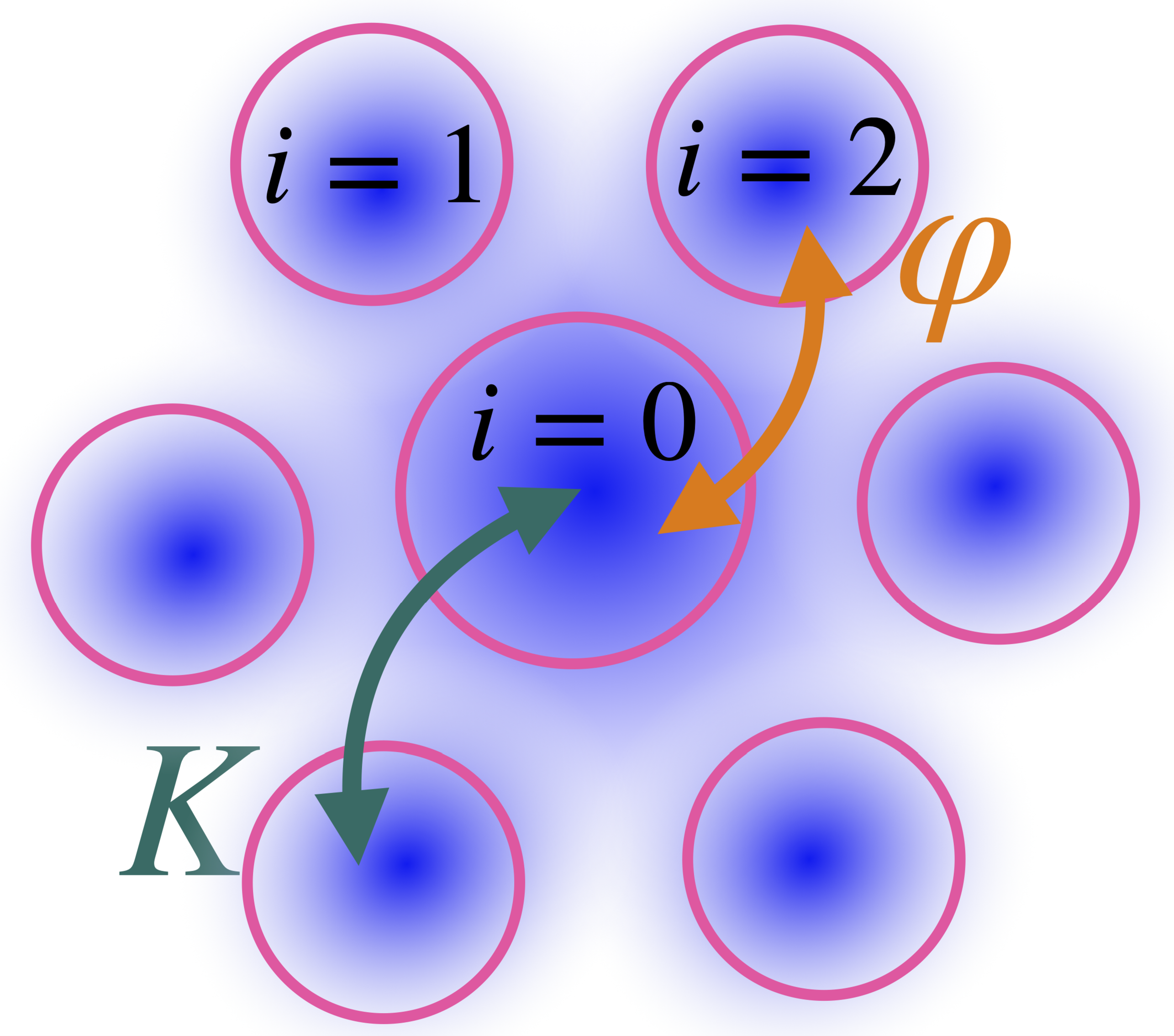}
 \caption{\label{fig:esq} Scheme of the Josephson-junctions
configuration in the $xy$ plane.  The central droplet ($i=0$) and the
six droplets forming the ring are coupled through the constant $K$,
with $\varphi$ representing the corresponding phase difference.  }
\end{figure}

\textit{System.}  We consider a triangular configuration comprising a
ring of droplets surrounding a central droplet, as schematically shown
in Fig. \ref{fig:esq}.  Such a configuration can be realized, for
instance, with a dipolar Bose gas composed of $N = 1.1 \times 10^{5}$
$^{162}$Dy atoms trapped by an axially symmetric potential with
frequencies $\{\omega_r, \omega_z\} = 2 \pi \times \{60, 120\}$ Hz,
with the dipoles aligned along the $z-$axis, as previously
investigated in Refs. \cite{gal20,aitor}. For $^{162}$Dy, the dipolar
scattering length is $a_{dd} = 130a_0$, and the $s$-wave scattering
length is set to $a_s = 92a_0$, where $a_0$ denotes the Bohr radius.
To create a population imbalance in the system, we introduce an
additional external potential composed of a set of Gaussian wells,
arranged with the same hexagonal symmetry as the ring of droplets,
which we will refer to as the \textit{egg-box} potential.  This
potential is used to prepare the initial state and is then removed,
allowing the system to evolve freely in the harmonic trap. By varying
selectively the intensity of the wells acting locally on the central
or ring droplets, we can prepare states with larger or smaller
imbalances relative to the equilibrium value. The same potential can
also be used to conveniently put the system under rotation, as
detailed in the Supplemental Material.

\textit{Asymmetric two-mode model.}  We consider that the central and
ring droplets are connected by Josephson junctions, with all external
droplets behaving identically due to the sixfold symmetry. That is,
all the droplets of the ring have the same population and phase at any
time, $N_i(t)=N_r(t)$ and $\phi_i(t)=\phi_r(t)$, where $1 \le i \le6$.
Hence, one can define only a single phase difference between the
centers of the central and the ring droplets
$\varphi(t)=\phi_0(t)-\phi_r(t)$, and a single imbalance given by
$Z(t)=(6 N_r(t)- N_0(t))/N = 1- 2 n_0(t)$, where $n_0(t)=N_0(t)/N$.
Here, we neglect the interchange of particles between the droplets and
the background density; that is, we consider $6N_r(t) + N_0(t)=N$.

The dynamics of the macroscopic variables $Z(t)$ and $\varphi(t)$ can
be described by the ATM model as presented in Ref. \cite{cat14}, whose
equations of motion can be simplified to
\begin{equation} \hbar \dot{Z}(t)= - K \sqrt{1-Z^2} \left( \frac{1-2
Z_e^2 + Z_e Z}{1- Z_e^2} \right) \sin{\varphi},
\label{zpun}
\end{equation}
\begin{equation} \hbar \dot{\varphi}(t)= U N (Z-Z_e ) + K \frac{(
Z-Z_e) (1+ 2 Z_e Z)}{ (1- Z_e^2) \sqrt{1-Z^2 }} \cos{\varphi},
\label{fipun}
\end{equation}
where $K$ and $U$ are related to the coupling and interaction energies
per particle, respectively (see Fig. \ref{fig:esq}).  The parameter
$Z_e$ represents the value of the imbalance at equilibrium, where
$\varphi=0$.  The above equations can be seen as the equations of
motion $\dot{\varphi}= {\partial H}/{\partial Z}$, $\dot{Z}=
-{\partial H}/{\partial \varphi}$ corresponding to the following
Hamiltonian,
\begin{align} H(Z,\varphi)=& \frac{N U }{2}( Z-Z_e)^2 \nonumber\\ -& K
\sqrt{1-Z^2} \left[\frac{1-2 Z_e^2 + Z_e Z}{1- Z_e^2} \right]
\cos(\varphi).
\label{hamiltoniano}
\end{align}
The dynamics of the system within the phase-space representation is
determined by the critical points of the Hamiltonian.  Such a set of
points consists of a minimum at $(Z_e,0)$, a saddle at $(Z_e, \pi)$,
and two maxima near $(|Z| \simeq 1, \pi) $. We recall that $(Z,\pi)$
and $(Z,-\pi)$ correspond to the same point in the phase space. Then,
the phase-space diagram $(Z,\varphi) $ exhibits Josephson oscillations
around the equilibrium point $(Z_e,0)$. Whereas the self-trapping
regime is separated from Josephson oscillations by two curves
$(Z_c,\varphi_c)$, called separatrices, that pass through the saddle
point, and can be numerically found by imposing $ H(Z_c,\varphi_c)=
H(Z_e,\pi) $. In particular, the value of the imbalances $Z_c^{\pm}$
for $\varphi_c=0$ can be roughly approximated by \cite{cat14},
\begin{equation} Z_c^{\pm} \simeq Z_e \pm \sqrt{ \frac{4 K
\sqrt{1-Z_e^2} }{U N}},
\label{zc}
\end{equation}
where the plus (minus) sign corresponds to the upper (lower) bound for
the Josephson oscillations.

\begin{figure*}[t]
\centerline{\includegraphics[width=\textwidth]{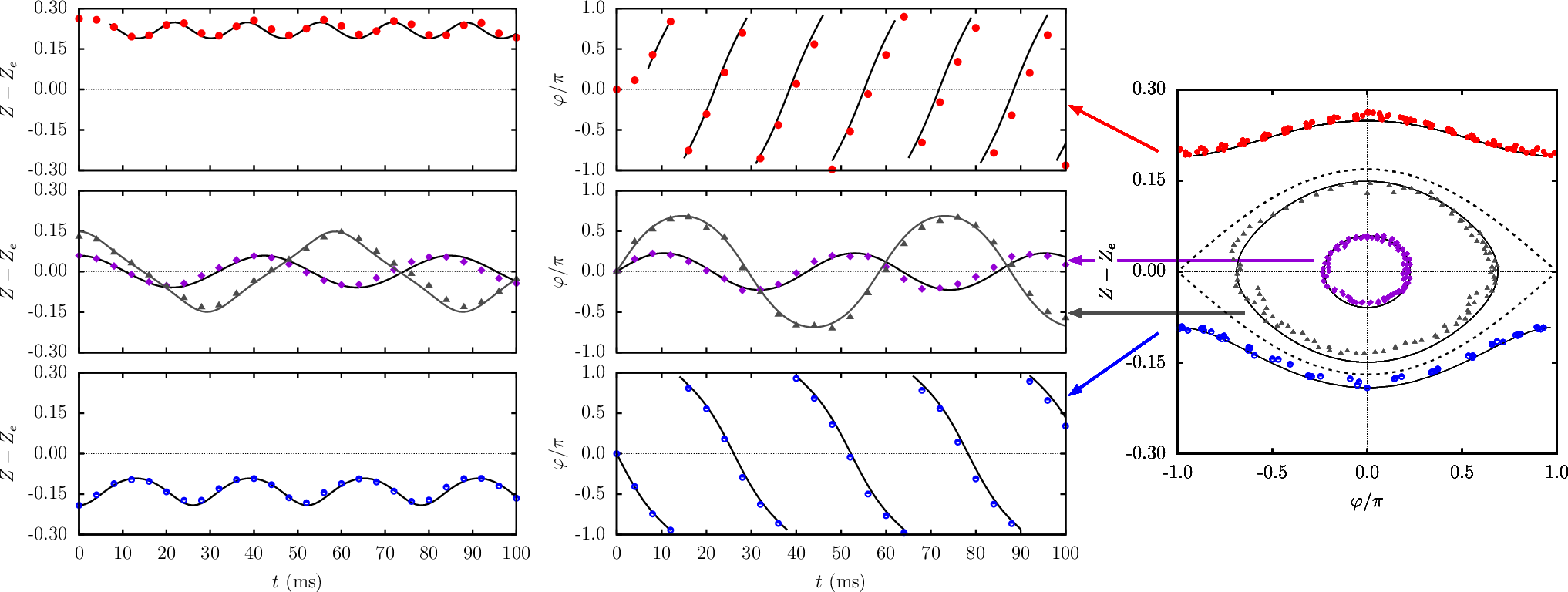}}
\caption{\label{fig:JS_ST} Josephson oscillations and self-trapping
  behavior in the non-rotating triangular supersolid, as obtained from
  the ATM model (solid black line) and GP simulations (colored
  dots). The leftmost panels display the evolution of the population
  imbalance $Z$ relative to the equilibrium value $Z_e\simeq 0.59$; the central
  panel, the evolution of the phase difference between the central and
  ring droplets (see Fig. \ref{fig:esq}); and the rightmost panel, the
  phase diagram in the ($Z$, $\varphi$) plane, with an additional
  trajectory represented by gray triangles.  The dashed line
  represents the separatrix between Josephson and self-trapping
  dynamics. All simulations were performed with $t_a=0$, except for
  the ST curve with the highest imbalance (red filled circles),  for
  which $t_a=10$ ms (see the Supplemental Material).}
\end{figure*}
\textit{Non-rotating system.}
The different dynamical behaviors predicted by the above model can be
explored through GP simulations (for the numerical details, which are
standard \cite{alana22,aitor}, see the Supplemental Material). The
system is prepared with different initial values of the imbalance,
$Z(0)$, and in the presence of the \textit{egg-box} potential, which
is completely removed at a time $t_a$.  The phases in the GP
simulations are calculated at the density maxima of the droplets. In
all cases, the initial phase difference between the central and ring
droplets is set to zero, $\varphi(0)=0$. Some of the numerical findings for
the non-rotating case are depicted in Fig. \ref{fig:JS_ST}.  By
initializing the system close to the equilibrium value $Z_e$, the
system reveals clear Josephson oscillations, similar to those reported
in Ref. \cite{Biagioni2024}, as shown in the central panels of the
figure. The phase oscillates around $\varphi = 0$, and the turning
points of $\varphi$ correspond to the points where $Z - Z_e = 0$.  By
increasing the initial offset $|Z-Z_e|$, we observe a self-trapping
behavior, with $Z$ oscillating without ever reaching $Z_e$.  This may
be seen for $Z> Z_c^+$ and $Z< Z_c^-$, top and bottom panels,
respectively.

In the rightmost panel of Fig. \ref{fig:JS_ST}, we present the phase
diagram in the $(Z, \varphi)$ plane, displaying some example
trajectories of Josephson oscillations and self-trapping dynamics,
corresponding to a couple of oscillation periods.  The numerical
results are compared with the predictions of the analytic ATM model,
shown as solid black lines. The parameters of the two-mode model, the
interaction energy $U$ and the coupling energy $K$, are extracted from
the output of the GP simulations, as detailed in the Supplemental
Material, yielding $K/\hbar \simeq 16$ Hz and $U/\hbar \simeq 0.016 $
Hz. The comparison shows remarkable agreement.

\begin{figure}
  \includegraphics[width=\columnwidth,clip=true]{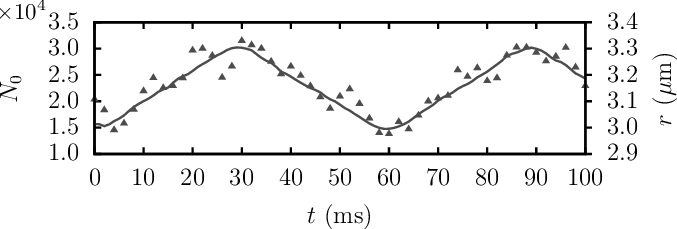}
  \caption{\label{fig:N0r} 
    Central droplet population $N_0$ (solid line, left-hand
      $y$-axis) and mean radius of the ring droplets $r$ (triangles,
      right-hand $y$-axis) as functions of time during the Josephson
      oscillation near the separatrix (see text). Both data are
      extracted from the GP simulations.}  
\end{figure}

In the middle panel of Fig. \ref{fig:JS_ST}, we show a comparison
between the \textit{small} and \textit{large} amplitude Josephson
oscillations, as displayed in the phase-space diagram. The former
corresponds to the inner trajectory (represented by purple diamonds),
while the latter corresponds to the outer trajectory near the
separatrix (represented by gray triangles).  
For the latter case, we show in Fig. \ref{fig:N0r} the value of $N_0$ 
and the position of the ring droplets $r$ (determined by the density maxima), 
as a function of time.
We note that in this case,
due to the large imbalance oscillation, the ring droplets also
exhibit significant movement around their equilibrium value,
in response to variations in the central droplet population, which
changes its size. Consequently the outer droplets should change their
positions.  Furthermore, due to the same fact, for the ST orbits
  the ring-droplets radius also perform oscillations, but with a
  smaller amplitude, around slightly lower (larger) values respect to
  the equilibrium one, in the case $Z(t) > Z_e$ ($Z(t) < Z_e$). Therefore, one could expect that the
position and characteristics of the junctions would
change, potentially leading to the breakdown of the model.  Moreover,
significant noise is observed, for which we cannot identify a clear
frequency. However, despite its simplicity, the ATM model correctly
predicts the character of Josephson oscillations near the separatrix,
where the evolution is very sensitive to parameters variations.  This
demonstrates the robustness of the model in describing the present
system across a wide range of configurations and provides further
evidence that the Josephson junction is the dominant mechanism driving
the dynamical evolution.

\textit{Rotating system.}  We now turn to consider the case of a
supersolid under rotation along the $z-$axis. To this end, we start
from a non-rotating supersolid and apply a torque using the
\textit{egg-box} potential, gradually increasing its angular velocity
$\Omega$ (see Supplemental Material for details).  As the system's
rotation increases, the equilibrium position $Z_e$ shifts, and the
distance between the droplets in the ring relative to the origin also
changes. These effects arise from a modification of the effective
trapping frequency, $\tilde{\omega} = \sqrt{\omega_r^2 - \Omega^2}$,
caused by the centrifugal force, as discussed further in the
Supplemental Material.  The \textit{egg-box} potential can be tuned so
that, when it is switched off, the relative imbalance
$Z - Z_e(\Omega)$ reaches the desired value.
After the potential is removed, the supersolid remains
self-sustaining, and the droplet positions and their relative
populations oscillate, leading to variations in the moment of inertia.
As a result, the rotation speed of the ring droplets adjusts
accordingly, with an amplitude variation below $10\%$ in the studied
cases, due to angular momentum conservation.

\begin{figure}[t] \centerline{\includegraphics[width=\columnwidth]{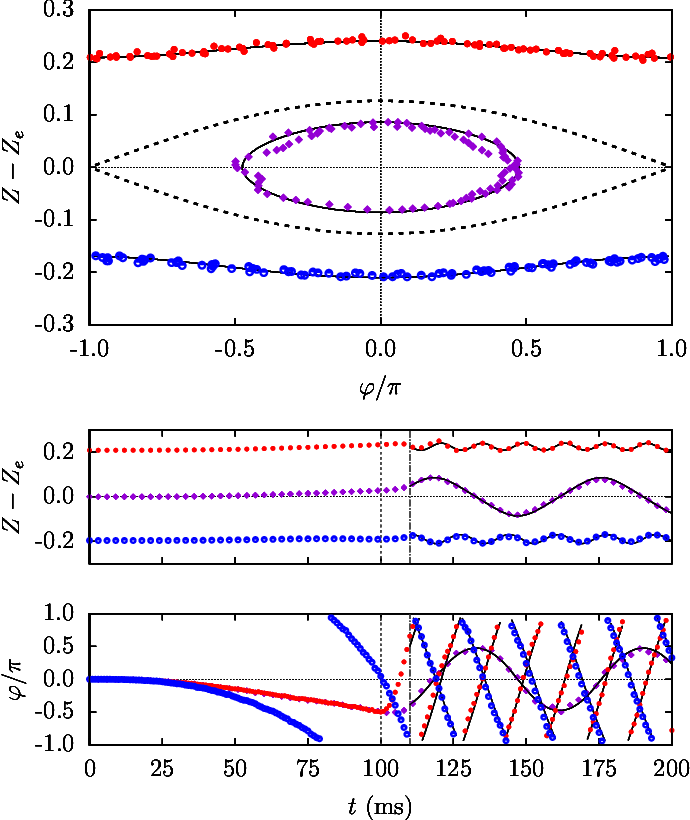}}
  \caption{\label{fig:rot_example} Phase-space portrait for dynamical
    simulations (top), evolution of the imbalance $Z$ (center), and
    evolution of the phase $\varphi$ (bottom), for a rotating
    supersolid with $\Omega=2\pi\times20$ Hz ($Z_e\simeq 0.59$). Each data point style
    refers to an evolution with different initial imbalance. In all
    cases the rotation has been implemented with a 100 ms linear ramp
    (from 0 to $\Omega$) followed by 10 ms relaxation of the
    \textit{egg-box} potential and 90 ms of free evolution,
    (encompassing various periods of the oscillation for all
    scenarios).  Black lines represent theoretical predictions based
    on the ATM model, similar to Fig. \ref{fig:JS_ST} (see there for a
    full description). The vertical dashed lines mark the end of the
    linear ramp and the end of the relaxation time $t_a=110$ ms, 
    where the applicability of the model starts. }
\end{figure}
\begin{figure}
  \includegraphics[width=\columnwidth,clip=true]{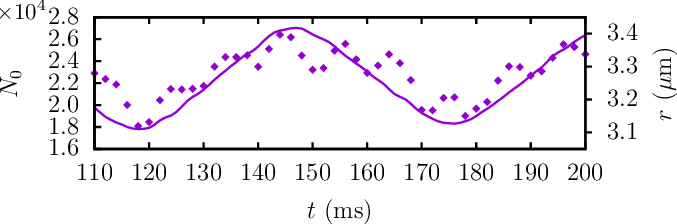}
  \caption{\label{fig:N0r_rot} Central droplet population $N_0$ and
    mean radius of the ring droplets $r$ as functions of time for the Josephson
    oscillation in the rotating supersolid (see text). The solid line
    marks $N_0$ (left-hand $y$-axis), while the diamonds correspond to
    $r$ (right-hand $y$-axis).}
\end{figure}

Nevertheless, despite the movement of the ring droplets in both the
radial and angular directions, and consequently the possible changes
in the structure and position of junctions, the coupling hopping seems
to describe once again all the orbits and, in particular, the orbit
near the separatrix.  These include Josephson oscillations and
self-trapping dynamics, observed for both $Z<Z_e$ and $Z>Z_e$, as
illustrated in Fig. \ref{fig:rot_example} for a representative case
with $\Omega=2\pi\times20$ Hz.
Similarly to the non-rotating case shown in Fig. \ref{fig:JS_ST}, this
figure presents typical trajectories of Josephson oscillations and
self-trapping dynamics in the phase-space representation (top panel),
along with the corresponding behavior of the imbalance and phase over
time (middle and lower panels, respectively).  In
Fig. \ref{fig:N0r_rot} we display a similar graph to that of
Fig. \ref{fig:N0r}, where we can also see that the variation in the
radial coordinate of the ring droplets is modulated by the central
population. However, here a faster superimposed oscillation can be
roughly identified.  Still, this extra frequency has no effect in the
macroscopic variables.  Remarkably, even in this case, the two-mode
model provides an accurate reference framework, with
$K/\hbar \simeq 10$ Hz and $U/\hbar \simeq 0.018$ Hz (see Supplemental
Material). Indeed, the model predictions (black lines) accurately
reproduce the dynamical behavior of the junction after the preparation
stage (the ramp) has been completed, as shown in the two lower panels,
and thus also the corresponding trajectories in the $(Z, \varphi)$
plane in the top panel.

\textit{Conclusions}.  We have demonstrated that a dipolar supersolid
can exhibit a self-trapping behavior besides the Josephson
oscillations observed in Refs. \cite{Biagioni2024,ilzhofer2021},
depending on the initial population imbalance.  For this we have
considered a triangular supersolid lattice with a central droplet
surrounded by six external droplets arranged in a
ring. Gross-Pitaevskii simulations revealed that introducing a
different number of particles in the central droplet leads to system
evolution that includes radial motion of the ring droplets, along with
angular velocity oscillations in the rotating case.  Our findings
extends the applicability of models beyond traditional systems, where
the sites and junctions are fixed by an external trap. Notably, we
have shown that the main features of this system can be accurately
reproduced by an asymmetric two-mode model, which depends on only two
parameters, $K$ and $U$, corresponding to the hopping between droplets
and the interaction energies per particle, respectively.  The
parameters correctly describe the macroscopic coordinate dynamics even
when large displacements of the ring droplets occur during the
evolution. Moreover, we find that the droplets oscillate around
different radius depending on the regime. In the case of the Josephson
regime, the oscillations occur around the equilibrium radius.  We have
investigated both the static configuration and the situation in which
the supersolid is set into rotation at almost constant angular
velocity $\Omega$, finding that the system's ability to display both
regimes, as well as the model description, remain robust even under
rotation where an angular oscillation of the ring droplets is added.
In addition, the ability to adjust the angular velocity offers an
interesting tool for tuning the strength of the weak link between the
droplets forming the supersolid.  We note that the centrifugal force
has a similar effect to changing the $s$-wave scattering length $a_s$
since in both cases the distances between droplets change.  Therefore,
we expect the model to remain valid across a wider set of parameters,
provided the geometric configuration of droplets is maintained.

These results provide a proof-of-concept that this geometry
offers a feasible and versatile experimental setup capable of
sustaining self-trapping regimes in supersolids in addition to
the recently observed Josephson oscillations \cite{Biagioni2024},
paving the way for future studies on their dynamics and stability
under various conditions.

\textit{Acknowledgments.}  We acknowledge
fruitful discussions with H. M. Cataldo.  This work was supported by
Grant PID2021-126273NB-I00 funded by MCIN/AEI/10.13039/501100011033
and by ``ERDF A way of making Europe'', by the Basque Government
through Grant No. IT1470-22, and by the European Research Council
through the Advanced Grant ``Supersolids'' (No. 101055319).
P.C. acknowledges support from CONICET and Universidad de Buenos
Aires, through grants PIP 11220210100821CO and UBACyT
20020220100069BA, respectively.

%

\end{document}


\title{Supplemental Material for ``Self-sustained Josephson dynamics and self-trapping in supersolids''}

\author{Aitor Ala\~na}
\affiliation{Department of Physics, University of the Basque Country UPV/EHU, 48080 Bilbao, Spain}
\affiliation{EHU Quantum Center, University of the Basque Country UPV/EHU, 48940 Leioa, Biscay, Spain}
\author{Michele Modugno}
\affiliation{Department of Physics, University of the Basque Country UPV/EHU, 48080 Bilbao, Spain}
\affiliation{IKERBASQUE, Basque Foundation for Science, 48009 Bilbao, Spain}
\affiliation{EHU Quantum Center, University of the Basque Country UPV/EHU, 48940 Leioa, Biscay, Spain}
\author{Pablo Capuzzi}
\affiliation{\hbox{Universidad de Buenos Aires, Facultad de Ciencias Exactas y Naturales, Departamento de Física, 1428 Buenos Aires, Argentina}}
\affiliation{CONICET - Universidad de Buenos Aires, Instituto de Física de Buenos Aires (IFIBA), 1428 Buenos Aires, Argentina}
\author{D. M. Jezek}
\affiliation{CONICET - Universidad de Buenos Aires, Instituto de Física de Buenos Aires (IFIBA), 1428 Buenos Aires, Argentina}

\date{\today}

\maketitle

\section{Gross-Pitaevskii simulations}

We describe the system using the usual extended Gross-Pitaevskii (GP)
theory, which includes the dipole-dipole interaction term
\cite{ronen2006} and quantum fluctuations in the form of the
Lee-Huang-Yang (LHY) correction, within the local density
approximation \cite{Fischer2006a,lima2012,wachtler2016,schmitt2016}.
The corresponding energy functional can be written as
$E[\psi,\psi^*]\equiv E_{\text{GP}}+E_{\text{dd}}+E_{\text{LHY}}$,
with
\begin{align}
E_{\text{GP}} &= 
\int \left[\frac{\hbar^2}{2m}|\nabla \psi(\bm{r})|^2 + V(\bm{r})n(\bm{r})+\frac{g}{2} n^2(\bm{r})
\right]d\bm{r}\,,
\nonumber\\
E_{\text{dd}} &=\frac{C_{\text{dd}}}{2}\iint n(\bm{r})V_{\text{dd}}(\bm{r}-\bm{r}')n(\bm{r}') d\bm{r}d\bm{r}'\,,
\label{eq:GPenergy}
\\
E_{\text{LHY}} &=\frac{2}{5}\gamma_{\text{LHY}}\int n^{5/2}(\bm{r})d\bm{r}\,.
\nonumber
\end{align}
The term
$E_{\text{GP}} = E_{\text{k}} + E_{\text{ho}} + E_{\text{int}}$
includes the kinetic, potential, and contact interaction energies,
where $V(\bm{r})$ denotes the external potential,
$n(\bm{r}) = |\psi(\bm{r})|^2$ the condensate density (normalized to
the total number of particles $N$), and $g = 4\pi\hbar^2 a_s/m$ the
contact interaction strength.  The term $E_{dd}$ represents the energy
associated with the dipole-dipole interaction, defined through the
potential $V_{\text{dd}}(\bm{r}) = (1 - 3\cos^2\theta) / (4\pi r^3)$,
where $C_{\text{dd}} \equiv \mu_0 \mu^2$ is the strength, $\mu$ is the
modulus of the dipole moment $\bm{\mu}$, and $\theta$ is the angle between the
dipole axis and the vector $\bm{r}$ representing the distance between
the dipoles, with $\cos\theta = \bm{\mu} \cdot \bm{r} / (\mu r)$.  The
LHY energy $E_{\text{LHY}}$ is fixed by the coefficient
$\gamma_{\text{LHY}}={128\sqrt{\pi}}{\hbar^{2}a_s^{5/2}}/(3m)\left(1+\epsilon_{\text{dd}}^{2}/2\right)$,
with $\epsilon_{\text{dd}}=\mu_0 \mu^2 N/(3g)$.

We employ numerical simulations both for computing the ground state
and the dynamical evolution.  The time-dependent generalized GP
equation is obtained as \cite{pitaevskii2016}
\begin{equation}\label{eq:gp}
    i\hbar{\partial_{t}\psi}={\delta E[\psi,\psi^*]}/{\delta \psi^*},
\end{equation}
where the energy functional $E[\psi,\psi^*]$ is the one defined above,
see Eq. (\ref{eq:GPenergy}).  For the ground state, we use the
minimization of the energy functional with the conjugate gradient
method (see, e.g., Refs. \cite{modugno2003,press07}). For the
dynamical evolution, we employ a FFT split-step algorithm
\cite{Jackson98}. The computations are performed on a box of
$24\mu$m$\times24\mu$m$\times24\mu$m, with a grid of
$192\times 192\times 64$ points.

\subsection{Preparation of population-imbalanced states}
\label{app:preparation}
The dipolar gas is trapped by an axially symmetric harmonic potential $V_{ho}(\bm{r})$, as described in the main text.
To create a population imbalance in the system and/or set the system
in rotation, we superimpose onto the harmonic trap an additional
  external potential, denoted as the \textit{egg-box} potential, for a
  time interval $t_a$. The additional potential consists of a set of
Gaussian wells arranged with the same hexagonal symmetry as the ring
of droplets. It can be written as follows,
\begin{equation}
\label{eq:egg}
\! V_{egg}(\bm{\rho},t)=V_0(t)e^{-2\rho^2\!/\sigma_{0}^2}{+V_{r}(t)\!\sum_{i=1}^{6}\!e^{-2|\bm{\rho}-\bm{\rho}_{0i}(t)|^2\!/\sigma_{0}^2}},
\end{equation}
with $\bm{\rho}$ representing the radial coordinate in the $(x,y)$
plane, $\bm{\rho}_{0i}$ the center-of-mass positions of the ring
droplets, $\sigma_{0}$ the widths of the Gaussians, and $V_0$ and
$V_r$ the strengths of the central and the ring wells, respectively.
Typical values that we employ are $\sigma_{0}=1.5\,\mu$m,
$V_{r}/h$ varied in the range $[62.5, 625]$ Hz, and $V_{0}/V_{r}$
in the range $[0.5,2.0]$.
The initial configuration is obtained as the ground state of $V(\bm{r})= V_{ho}(\bm{r}) + V_{egg}(\bm{\rho},t=0)$.
This potential provides an attractive force to control the position
and population of the droplets. By tuning the relative strength of the
central and ring wells, one can adjust the initial imbalance of $Z$,
either above or below the equilibrium value $Z_{e}$. Dynamically
changing the values of $\bm{\rho}_{0i}(t)$ corresponds to imprinting a
torque onto the system, making it rotate. Specifically, we consider
rotations at variable angular velocity $\Omega(t)$, which can be
obtained by setting
$\bm{\rho}_{0i}(t) = \hat{R}[\theta(t)]\bm{\rho}_{0i}(0)$, where
$\dot{\theta}(t)=\Omega(t)$ and $\hat{R}$ is the rotation
operator. Notably, the potential in Eq. \eqref{eq:egg} also provides a
trapping force, preventing the outward drift of the droplets.  In both
the rotating and non-rotating cases, the removal of the
\textit{egg-box} potential at $t_a$ marks the point when the model
becomes applicable, as the system evolves freely in the harmonic trap.

\textit{Non-rotating system.}  In this case, we use a static potential
$V_{egg}(\bm{\rho})$.  For a population fraction of the central
droplet larger (smaller) than the equilibrium value, with $Z < Z_e$
($Z > Z_e$), we turn on $V_0$ ($V_r$), while keeping the other
potential set to zero. Such a potential can be tuned to achieve the
desired initial value of the imbalance, $Z(0)$.  The potential is then
switched off at $t=t_a$.   The simulations shown in Fig. 2
  correspond to: (top panel) $Z(0) \simeq 0.85$; (middle panel,
  colored diamonds) $Z(0)\simeq 0.64$; and (middle panel, colored
  triangles) $Z(0)\simeq 0.71$; (bottom panel) $Z(0)\simeq 0.40$.
We note that the case of the highest initial imbalance, 
$Z(0) \simeq 0.85$, required a nonzero $t_a$ to avoid excitations
favoring a different geometric configuration, as the central droplet
is highly depleted in this case.

\textit{Rotating system.}  Here, we use a fully time-dependent
potential, $V_{\text{egg}}(\bm{\rho}, t)$.  The initial imbalance at
$t = 0$ is set in the same way as in the previous non-rotating case.
The lattice structure is then rotated rigidly by periodically driving
the positions $\bm{\rho}_{0i}(t)$, as described earlier, by increasing
the rotation frequency from $0$ to $\Omega(t_f)$ via a linear ramp
over a time $t_f = 100$ ms.  At the end of this stage, the
\textit{egg-box} potential is linearly ramped down to zero over $10$
ms, by changing the values of both $V_r(t)$ and $V_0(t)$.

\begin{figure}[t]
 \includegraphics[width=\columnwidth]{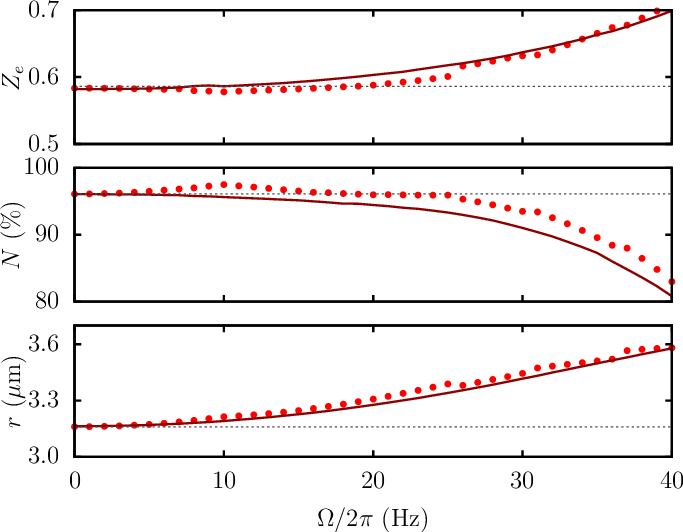}%
 \caption{\label{fig:Z_rot} Equilibrium imbalance, total populations
   of the droplets (relative to the maximum population), and the mean
   radius of the ring at the equilibrium states as functions of
   $\Omega$.  Solid lines correspond to the non-rotating system with
   $ \tilde{\omega}= \sqrt{\omega_r^2 - \Omega^2}$, and red dots are
   obtained from actual ground state simulations in the rotating
   scenario. }
\end{figure}

It is also worth noting that, when the system is rotating, the
equilibrium configuration changes, affecting both the equilibrium
position $Z_e$ and the positions of the ring droplets. This can be
attributed to the centrifugal force, which effectively weakens the
radial trapping potential, as illustrated in
Fig. \ref{fig:Z_rot}. Owing to this, for a given rotation frequency,
we also adjust the radius $|\bm{\rho}_{0i}(0)|$ of the ring droplets
to set the desired initial conditions for the two-mode model in each
simulation.

%
\section{ Parameters of the ATM model }
\label{app:parameters}

The parameters of the two-mode model, namely the interaction energy
$U$ and coupling energy $K$, can be extracted from the output of the
GP simulations as follows.

In the \textit{non-rotating} case, the parameter $U$ can be determined
from the ST dynamics in the bottom panels in Fig. 2 of the main
article (blue empty circles) while disregarding the second term in
Eq. (2) of the main text, assuming $K \ll NU$. This leads to
\begin{equation}
 U/\hbar \simeq \frac{2\pi}{ T N |\bar{Z}-Z_e|},
\label{ug}
\end{equation}
where $ \bar{Z}$ is the mean value in an oscillation and $T$ is its
period.  From the simulation, we have $T \simeq 25$ ms,
$\bar{Z} \simeq 0.45$, yielding $U/\hbar \simeq 0.016 $ Hz (recalling
that $N = 1.1 \times 10^5$).

The coupling energy $K$ can be instead obtained from the Josephson
oscillation. Indeed, in the small amplitude regime, the oscillation
period $T_{J}$ can be written as
\begin{equation}
 T_{J} \simeq \frac{  2 \pi\hbar}{ \sqrt{ K U N \sqrt{1-Z_e^2} } },
\end{equation}
so that
\begin{equation}
 K/\hbar \simeq \left[\frac{2\pi}{ T_{J}}\right]^2 \frac{ 1}{ U N \sqrt{1-Z_e^2} }.
\end{equation}
From the Josephson oscillations in the central panels of Fig. 2 of the
main article (magenta diamonds), one obtains $T_J \simeq 41.1$
ms. Using the above equation, we can then estimate $K/\hbar \simeq 16$
Hz.  The above results also allow us to check the validity of the
assumption $K \ll NU$, which in the present case reads
$16 \ll 1.76 \times 10^{3}$. This confirms the consistency of the
approach.

In the \textit{rotating case}, the same approach can be used, this
time determining the parameters $U$ and $K$ from the simulations shown
in Fig. 4 of the main article.  Specifically, $U$ is derived from the
self-trapping oscillation with $\bar{Z} - Z_e \sim 0.2$ (red squares
in the figure), yielding $U/\hbar \simeq 0.018\, \mathrm{Hz}$.  The
value of $K$ can instead be extracted from the period of the small
amplitude Josephson oscillations, by discarding a transient initial
stage in which there are noisy oscillations produced by the ramp.
From this, we estimate $T_J \simeq 49 \, \mathrm{ms}$, yielding
$K/\hbar \simeq 10 \, \mathrm{Hz}$.

%